\documentclass[11pt]{article}
\usepackage{amsmath,amsthm,amsfonts,amssymb, graphicx,mathabx,epsfig}
\usepackage{bbm}
\usepackage{bm}
\usepackage{authblk}
\usepackage{setspace}
\usepackage[usenames]{color}
\usepackage[active]{srcltx}

\usepackage{graphicx,color}
\usepackage{footmisc}
\usepackage{amsmath}
\usepackage{amsfonts}
\usepackage{amssymb}
\usepackage{bbm}
\usepackage{epstopdf}
\usepackage{color}
\usepackage{dsfont}
\numberwithin{equation}{section}
\usepackage{tikz}
\usepackage{pgfplots}
\usepackage{subfig}
\usetikzlibrary{fit}
\usetikzlibrary{shapes.geometric}
\usetikzlibrary{decorations.pathmorphing}
\usepackage{slashed}
\usepackage{color}

\usepackage{amsmath}
\usepackage{slashed}

\usepackage{amsmath,amsthm,amsfonts,amssymb, graphicx,mathabx,epsfig}
\usepackage{bbm}
\usepackage{authblk}
\usepackage{bm}
\usepackage{graphicx,color}
\usepackage{footmisc}
\usepackage{epstopdf}
\usepackage{dsfont}
\usepackage{pgfplots}
\usepackage{subfig}
\usetikzlibrary{fit}
\usepackage{slashed}

\usepackage{graphicx}
\usepackage{dcolumn}
\usepackage{bm}
\usepackage{epstopdf}

\newcount\driver
\newcount\bozza

scaled\magstep1                  
scaled\magstep1 \font\msytw=msbm10 scaled\magstep1
 
scaled\magstep1                 \font\indbf=cmbx10 scaled\magstep2

scaled \magstep2

{\count255=\time\divide\count255 by 60
\xdef\hourmin{\number\count255}
        \multiply\count255 by-60\advance\count255 by\time
   \xdef\hourmin{\hourmin:\ifnum\count255<10 0\fi\the\count255}}

\let\a=\alpha \let\b=\beta    \let\g=\gamma     \let\d=\delta     \let\e=\varepsilon
  \let\h=\eta           \let\l=\lambda
\let\m=\mu    \let\n=\nu      \let\x=\xi        \let\p=\pi        \let\r=\rho
\let\s=\sigma \let\t=\tau            
\let\ps=\psi   \let\o=\omega     
 \let\D=\Delta       \let\L=\Lambda    
             
\let\O=\Omega

\def\EE{{\cal E}}\def\VV{{\cal V}}

\def\RR{{\cal R}}

\def\pp{{\bf p}}\def\xx{{\bf x}}
  
\def\yy{{\bf y}}\def\kk{{\bf k}}\def\nn{{\bf n}}

       \def\oo{{\underline \omega}}
\def\ee{{\underline \varepsilon}}

\def\RRR{\hbox{\msytw R}}

        \def\ZZZ{\hbox{\msytw Z}}

        \def\EE{\hbox{\msytw E}}

\let\==\equiv

\let\io=\infty
\let\0=\noindent

\def\*{{\hfill\break\null\hfill\break}}

\def\media#1{{\langle#1\rangle}}

\def\tilde#1{{\widetilde #1}}

\def\tende#1{\,\vtop{\ialign{##\crcr\rightarrowfill\crcr
             \noalign{\kern-1pt\nointerlineskip}
             \hskip3.pt${\scriptstyle #1}$\hskip3.pt\crcr}}\,}
\def\otto{\,{\kern-1.truept\leftarrow\kern-5.truept\to\kern-1.truept}\,}

\def\wh#1{\widehat{#1}}
\def\hat#1{\wh{#1}}
\def\sqt[#1]#2{\root #1\of {#2}}

\def\bp{{\bar \ps}}

\def\EE{{\cal E}}\def\VV{{\cal V}}

\def\RR{{\cal R}}

\def\T#1{{#1_{\kern-3pt\lower7pt\hbox{$\widetilde{}$}}\kern3pt}}
\def\VVV#1{{\underline #1}_{\kern-3pt
\lower7pt\hbox{$\widetilde{}$}}\kern3pt\,}
\def\W#1{#1_{\kern-3pt\lower7.5pt\hbox{$\widetilde{}$}}\kern2pt\,}

\def\indica{\leaders \hbox to 0.5cm{\hss.\hss}\hfill}
\def\guida{\leaders\hbox to 1em{\hss.\hss}\hfill}
\mathchardef\oo= "0521

\def\pp{{\bf p}}\def\xx{{\bf x}}
\def\yy{{\bf y}}\def\kk{{\bf k}}\def\nn{{\bf n}}

\def\oo{{\underline \omega}}

\def\qed{\raise1pt\hbox{\vrule height5pt width5pt depth0pt}}
 
  \def\bp{{\bar p}} 

\def\indic{\hbox{\raise-2pt \hbox{\indbf 1}}}

\def\RRR{\hbox{\msytw R}}

 \def\ZZZ{\hbox{\msytw Z}}

\def\ins#1#2#3{\vbox to0pt{\kern-#2 \hbox{\kern#1 #3}\vss}\nointerlineskip}

\newdimen\xshift \newdimen\xwidth \newdimen\yshift

\def\insertplot#1#2#3#4#5#6{%
\xwidth=#1pt \xshift=\hsize \advance\xshift by-\xwidth \divide\xshift by 2%
\begin{figure}[ht]
\vspace{#2pt} \hspace{\xshift}
\begin{minipage}{#1pt}
#3 \ifnum\driver=1 \griglia=#6
\ifnum\griglia=1 \openout13=griglia.ps \write13{gsave .2
setlinewidth} \write13{0 10 #1 {dup 0 moveto #2 lineto } for}
\write13{0 10 #2 {dup 0 exch moveto #1 exch lineto } for}
\write13{stroke} \write13{.5 setlinewidth} \write13{0 50 #1 {dup 0
moveto #2 lineto } for} \write13{0 50 #2 {dup 0 exch moveto #1
exch lineto } for} \write13{stroke grestore} \closeout13
\includegraphics{griglia.ps} \fi
\includegraphics{#4.ps}\fi%
\ifnum\driver=2 \fi
\end{minipage}
\caption{#5}
\end{figure}
}

\newdimen\shift \shift=-1.5truecm
\def\lb#1{%
\ifnum\bozza=1
\label{#1}\rlap{\hbox{\hskip\shift$\scriptstyle#1$}}
\else\label{#1} \fi}

\def\be{\begin{equation}}
\def\ee{\end{equation}}
\def\bea{\begin{eqnarray}}\def\eea{\end{eqnarray}}
\def\bean{\begin{eqnarray*}}\def\eean{\end{eqnarray*}}
\def\bfr{\begin{flushright}}\def\efr{\end{flushright}}
\def\bc{\begin{center}}\def\ec{\end{center}}
\def\bal{\begin{align}}\def\eal{\end{align}}
\def\ba#1{\begin{array}{#1}} \def\ea{\end{array}}
\def\bd{\begin{description}}\def\ed{\end{description}}
\def\bv{\begin{verbatim}}\def\ev{\end{verbatim}}
\def\nn{\nonumber}
\def\Halmos{\hfill\vrule height10pt width4pt depth2pt \par\hbox to \hsize{}}
\def\pref#1{(\ref{#1})}

\def\ins#1#2#3{\vbox to0pt{\kern-#2 \hbox{\kern#1 #3}\vss}\nointerlineskip}

\newdimen\xshift \newdimen\xwidth \newdimen\yshift
\newcount\griglia

\def\insertplot#1#2#3#4#5#6{%
\xwidth=#1pt \xshift=\hsize \advance\xshift by-\xwidth \divide\xshift by 2%
\begin{figure}[ht]
\vspace{#2pt} \hspace{\xshift}
\begin{minipage}{#1pt}
#3 \ifnum\driver=1 \griglia=#6
\ifnum\griglia=1 \openout13=griglia.ps \write13{gsave .2
setlinewidth} \write13{0 10 #1 {dup 0 moveto #2 lineto } for}
\write13{0 10 #2 {dup 0 exch moveto #1 exch lineto } for}
\write13{stroke} \write13{.5 setlinewidth} \write13{0 50 #1 {dup 0
moveto #2 lineto } for} \write13{0 50 #2 {dup 0 exch moveto #1
exch lineto } for} \write13{stroke grestore} \closeout13
\includegraphics{griglia.ps} \fi
\includegraphics{#4.ps}\fi%
\ifnum\driver=2 \fi
\end{minipage}
\caption{#5}
\end{figure}
}

\newdimen\shift \shift=-1.5truecm
\def\lb#1{%
\label{#1}\rlap{\hbox{\hskip\shift$\scriptstyle#1$}}
\else\label{#1} \fi}

\def\be{\begin{equation}}
\def\ee{\end{equation}}
\def\bea{\begin{eqnarray}}\def\eea{\end{eqnarray}}
\def\bean{\begin{eqnarray*}}\def\eean{\end{eqnarray*}}
\def\bfr{\begin{flushright}}\def\efr{\end{flushright}}
\def\bc{\begin{center}}\def\ec{\end{center}}
\def\bal{\begin{align}}\def\eal{\end{align}}
\def\ba#1{\begin{array}{#1}} \def\ea{\end{array}}
\def\bd{\begin{description}}\def\ed{\end{description}}
\def\bv{\begin{verbatim}}\def\ev{\end{verbatim}}
\def\nn{\nonumber}
\def\Halmos{\hfill\vrule height10pt width4pt depth2pt \par\hbox to \hsize{}}
\def\pref#1{(\ref{#1})}

\usepackage{amsmath}
\usepackage{amsfonts}
\usepackage{amssymb}
\usepackage{epstopdf}

\font\msytw=msbm9 scaled\magstep1 
scaled\magstep1

\let\a=\alpha \let\b=\beta  \let\g=\gamma  \let\d=\delta
\let\e=\varepsilon
  \let\h=\eta     \let\l=\lambda
\let\m=\mu    \let\n=\nu    \let\x=\xi     \let\p=\pi    \let\r=\rho
\let\s=\sigma \let\t=\tau    
\let\ps=\Psi   \let\o=\omega
 \let\D=\Delta  \let\L=\Lambda 
         
\let\O=\Omega 

\def\EE{{\cal E}} \def\VV{{\cal V}}

\def\RR{{\cal R}}

 \def\pp{{\bf p}}
 \def\xx{{\bf x}} \def\yy{{\bf y}} 
\def\kk{{\bf k}}

\def\nn{\nonumber}

\def\RRR{\hbox{\msytw R}}

 \def\ZZZ{\hbox{\msytw Z}}

\def\\{\hfill\break}
\def\={:=}
\let\io=\infty
\let\0=\noindent
\def\media#1{{\langle#1\rangle}}

\def\tende#1{\,\vtop{\ialign{##\crcr\rightarrowfill\crcr\noalign{\kern-1pt
    \nointerlineskip} \hskip3.pt${\scriptstyle #1}$\hskip3.pt\crcr}}\,}
\def\otto{\,{\kern-1.truept\leftarrow\kern-5.truept\to\kern-1.truept}\,}

\def\wh{\widehat}
\def\to{\rightarrow}

\def\qed{\hfill\raise1pt\hbox{\vrule height5pt width5pt depth0pt}}

\def\be{\begin{equation}}
\def\ee{\end{equation}}
\def\bp{\begin{pmatrix}}
\def\ep{\end{pmatrix}}
\def\bea{\begin{eqnarray}}
\def\eea{\end{eqnarray}}
\def\nn{\nonumber}
\def\pref#1{(\ref{#1})}

\def\lb{\label}

\begin{document}

\title{Universality for critical lines for Ising, Vertex and Dimer models}

\author[1]{Vieri Mastropietro}
\affil[1]{University of Milano, Department of Mathematics ``F. Enriquez'', Via C. Saldini 50, 20133 Milano, Italy}

\maketitle
\begin{abstract} In planar lattice statistical mechanics models like coupled Ising with quartic interactions, vertex and dimer models, the exponents depend
on all the Hamiltonian details. This corresponds, in the Renormalization Group language,
to a line of fixed points. A form of universality is expected to hold, implying that
all the exponents can be expressed by exact "Kadanoff" relations in terms of a single one of them. 
This conjecture has been recently established and we review
here the key step of the proof,
obtained by rigorous Renormalization Group methods and valid irrespectively on the 
solvability of the model. The exponents are expressed by
convergent series in the coupling and, thanks to a set of cancellations due to emerging chiral  symmetries, the extended scaling relations are proven to be true.
\end{abstract}

\section{Introduction}

It is a matter of fact that several complex systems show a remarkable 
{\it universal} behavior in some physical observables.
Let us consider for instance the $d=2$ Ising model, which provides a description of
planar magnetic materials; in particular an atomic layer of ferromagnetic
iron deposited on a non-magnetic substrate \cite{B1}
or layers of gold \cite{B2}. The Ising model can be exactly solved \cite{O}
and 
{\it critical exponents} can be computed, like the one of magnetization $\b=1/8=0.125$ and of the correlation length  
$\n=1$.
Experiments show a remarkable
agreement with such values: actually
$\b=0,13\pm 0,02$ \cite{B1} or 
$\b= 0.13\pm 0.022$ and $\n=1.02\pm 0.02$ \cite{B2}. 
The two materials have a different microscopic structure, which
depends on a number of parameters like atomic weights and crystalline structure. 
One can a priori expect that their properties
are {\it qualitatively} similar, but this {\it quantitative}
agreement  needs an explanation.
It is also remarkable the agreement with the Ising model, which indeed provides a very simplified 
description of such systems.
More quantitatively
the mechanism of universality in such systems can be described in the following way, see e.g. \cite{Ba}. We can imagine that they are driven by an Hamiltonian of the form
\be
H=H_0+\l H_1\label{c6}
\ee
where $H_0$ is the Hamiltonian of some simple model, like the Ising model,
and $H_1$ contains all the microscopic details. Then universality in the above cases means that the exponents are $\l$- independent, that is $X(\l)=X(0)$ if $X$ is an exponent. 

There are however also cases in which the exponents are functions of the parameters, as observed in certain materials \cite{B3}. This is indeed a rather common situation
in planar lattice systems. In these cases the exponents of different systems cannot coincide, but Kadanoff proposed a notion of universality true in this case, consisting in the fact that {\it all
the exponents can be expressed in terms of a single one}; this notion 
was later extended by Haldane \cite{Ha}.
Even if the exponents depend on all microscopic details, the knowledge of one
of them (by exact computation or measurement) allows the exact knowledge of all 
the others.

In the above systems the underlying physics
is essentially classical  but universality is present also in low temperature experiments,
where the microscopic description is based on quantum mechanics. 
Examples are in the transport properties of electrons, like
the optical conductivity in {\it graphene}, which is given with very high precision
by $\s={\pi\over 2}{e^2\over h}$,
where $e$ is the electric charge and $h$ is the Planck constant, and is independent from all the material microscopic details \cite{B4}; or in the exact quantization of transport coefficients
in topological Hall insulators.
%

In order to explain the universality phenomenon, one needs to start from microscopic Hamiltonian models and compute thermodynamical quantities according to the axioms of statistical mechanics. 
A powerful approach is provided by the
{\it Renormalization Group} (RG) of Wilson \cite{W}, in which systems belong to universality classes 
associated to fixed points.
The classification distinguishes between a) 
gaussian fixed  points, in which the behavior is the free or the mean field one; b) non trivial fixed points distant from the gaussian one, a celebrated example being provided
by the three dimensional  Ising model; c) lines of fixed points, to which are associated continuously varying exponents. In most physical applications the fixed points are studied under truncations and approximations, but RG ideas can be also used in rigorous proofs.
In particular, a rigorous understanding of universality has been reached in the case of gaussian fixed point. In contrast, the case of  non trivial fixed points require the introduction
"by hand" of a small parameter (like in the $\e$-expansion) to make such point close to the gaussian one, but in models with physical relevance (as the 3D Ising model)
such a parameter is lacking.
Lines of fixed points are somewhat intermediate as they
can be 
close to the gaussian one for certain region of parameters.
Nevertheless, they have a sort of a non-perturbative nature
as their existence requires compensation at any order in the expansions. 

The natural context for understanding universality in presence of lines of fixed 
points are planar statistical mechanics models. Following \cite{Ka}, universality
means that critical exponents are connected by a number of extended scaling relations, implying that all
the exponents can be expressed in terms of a single one of them.
Such a conjecture has been verified in certain special exactly solvable models
\cite{Ha}, \cite{LP}. There were attempts 
to prove such relations in general by relating planar lattice models
to gaussian free fields in the continuum, with a suitable correspondence of lattice
correlations with averages of function of the gaussian field, see  e.g.
\cite{K11}, \cite{K12}, \cite{K13}. This mapping is however not rigorous, except 
in the scaling limit for the planar Ising  \cite{D} or dimer model \cite{kos}, which do not belong to this class.

A rigorous approach based on Renormalization Group have finally proved
universality 
in the case of lines of fixed points
for planar lattice statistical mechanics models,
like coupled Ising, dimer or vertex models. The proof holds for both solvable and not solvable models. The exponents are expressed by
convergent series in the coupling and, thanks to a set of cancellations due to emerging chiral  symmetries, the extended scaling relations are proven to be true.
We review the main steps of the proof
of universality in such models, as achieved in \cite{M1},
\cite{BM}, \cite{GM},\cite{BFM},\cite{M2}, \cite{GMT1}, \cite{GMT2}.
This notion of universality is indeed much more general and indeed
by similar methods one can prove also universality 
in condensed matter models, like Luttinger liquids \cite{BFM3},\cite{BFM4} and Hall insulators \cite{AMP}, but such results will be not reviewed here.

\section{Planar lattice models}

The Hamiltonian of the nearest-neighbor planar Ising model is 
\be
H_J=-J\sum_{\xx\in\L} (\s_{\xx}\s_{\xx+{\bf e}_0}+\s_{\xx}\s_{\xx+{\bf e}_1})\label{is}
\ee
where $\L\subset \ZZZ^2$ is a square box of side $L$ with periodic boundary conditions,
$\s_\xx=\pm$ are spin variables , ${\bf
e}_0=(1,0),{\bf e}_1=(0,1)$. The Hamiltonian describes a system of dipoles which can be imagined as arrows pointing up $\s=+$ or down $\s=-$, and there is a contribution to the energy $-J$
if two nearest-neighbors dipoles point in the same direction, and $J$ in the opposite case. The factor $J$, the exchange energy, is the only parameter entering in the Hamiltonian.
The model describes a magnet but it
contains a number of drastic simplifications. First of all, there is no reason why  the spins should point only in two directions.
In addition, 
in general not only nearest neighbor spin should interact.

To address part of such limitations one can introduce a
generalized {\it
Ising model} in $d=2$ 
\bea
&&H=H_J+\l H_1\nn\\
&&H_1=
\sum_{j=0,1}\sum_{\xx,\yy\in\L} v_j(\xx-\yy)\s_\xx\s_{\xx+{\bf
e}_j} \s_\yy\s_{\yy+{\bf e}_j}\label{c7}
\eea
where $v_j(\xx-\yy)$ can be assumed short ranged. In the particular case  $v_j(\xx-\yy)=\d_ {\yy,\xx+{\bf
e}_j}$ one gets 
\be
H_1=
\sum_{j=0,1}\sum_{\xx\in\L} (\s_\xx \s_{\xx+2 {\bf e}_0}+\s_\xx \s_{\xx+2 {\bf e}_1})
\ee
describing the next to nearest neighbor interaction. 

Certain materials are described by
two Ising models with a quartic term with coupling $\l$; the Hamiltonian is  
\be H(\s,\s')=H_J(\s) +
H_{J'}(\s')- \l V(\s,\s')-J_4 \label{c11}\ee 
with $J_4$ a constant and
\be
V=
\sum_{j=0,1}\sum_{\xx,\yy \in\L} v_j(\xx-\yy)\s_\xx\s_{\xx+{\bf
e}_j} \s'_\xx\s'_{\xx+{\bf e}_j} 
\ee
with $v_j(\xx-\yy)$ is a short range potential. Two special choices are remarkable.

With the choice
\be
V=\sum_{j=0,1}\sum_{\xx \in\L} \s_\xx\s_{\xx+{\bf
e}_j} \s'_\xx\s'_{\xx+{\bf e}_j} \label{c2}\ee
the model can be mapped in the
{\it Ashkin-Teller model} with a suitable choices of the parameters, see \S 12.9 of \cite{Ba}. Such a model can be considered a generalization of the Ising model in which the spin assumes four values,
denoted by $A,B,C,D$. The energy is $\sum_{i=1}^4 n_i \e_i$ where the energy  $e_0$ correspond to $AA,BB,CC,DD$; $\e_1$ to $AB,CD$; $\e_2$ to $AC,BD$;
$\e_3$ for $AD,BC$. The energy can be written as the Hamiltonian \pref{c11}, \pref{c2} with the identification
\bea
&&J=-(\e_0+\e_1-\e_3-\e_4)/4 \quad J'=-(\e_0+\e_2-\e_3-\e_1)/4\nn\\
&&\l=-(\e_0+\e_3-\e_1-\e_2)/4\quad J_4=-(\e_0+\e_1+\e_3+\e_4)/4
\eea
Another important choice is
\be
V=\sum_{j=0,1}\sum_{\xx \in\L} \s_{\xx+{\bf
e}_0}\s_{\xx+{\bf e}_0+{\bf
e}_1} \s'_{\xx+{\bf
e}_1}\s'_{\xx+{\bf
e}_0+{\bf
e}_1}\label{111}
\ee
and if $J=J'$ is
equivalent to {\it eight vertex} model, with identification of parameters  see \S 10.3 of \cite{Ba}
\be
a=e^{2 J+\l}\quad\quad  b=e^ {-2 J+\l} \quad c=d=e^{-\l}\label{c20} 
\ee
where $a,b,c,d$ are the weight associates to the arrows. 

The thermodynamical averages are defined as, given $A=A(\underline\s)$, $\underline\s= 
\{\s_\xx\}_{\xx\in\L}$,
$\O=\{\pm 1\}^\L$,
\be
<A>_{L,\b}={\sum_{\underline\s\in \O}e^{-\b H(\underline\s)} A\over\sum_{\underline\s\in \O}e^{-\b H(\underline\s)}}
\ee
where $\b$ is the inverse temperature
and we define $<A>_{\b}=\lim_{L\to\io}<A>_{L,\b}$. 

Important correlations in the case of the generalized Ising model are
 the {\it spin} correlations $<\s_\xx\s_\yy>$
and the {\it energy} correlation $<\r_\xx\r_\yy>_L$ where
$\r_\xx=\s_\xx\s_{\xx+{\bf e}_0}+\s_\xx\s_{\xx+{\bf e}_1}
$; the truncated correlation is 
$<\r_\xx;\r_\yy>_T=<\r_\xx\r_\yy>-<\r_\xx><\r_\yy>$. 
Close to a phase transition the correlations are expected to behave as, if $t=O(\b-\b_c)$
\be
<\s_\xx\s_\yy>_\b\sim {e^{- \x^{-1}  |\xx-\yy|}\over |\xx-\yy|^\h}\quad\quad 
<\r_\xx\r_\yy>_{T,\b}\sim {e^{- \x^{-1}  |\xx-\yy|}\over |\xx-\yy|^{2 X_e}}
\quad\quad \x=t^{-\n}\label{sap}
\ee
whee $\h,\n,X$ are critical exponents and $1/\b_c$ is the critical temperature; at $\b=\b_c$
the correlations decay as a power law. 

In the case of coupled Ising models \pref{c11}, one can introduce respectively the {\it polarization},
the {\it energy} and the {\it crossover
}  correlations, and one expects 
at the critical temperatures $\b_c^i$ (there are in general two critical temperatures except when $J\not =J'$) there is a power law decay.
\bea
&&<\s_\xx\s'_\xx;\s_\yy\s'_\yy>_{T,\b}\sim  {1\over |\xx-\yy|^{2 X_{P} }}\\
&& <(\r_\xx+\r'_\xx)(\r_\yy+\r'_\yy)>_{T,\b}\sim {1\over |\xx-\yy|^
{2 X_{e} }}
\quad 
<(\r_\xx-\r'_\xx); (\r_\yy-\r'_\yy)>_{T,\b}\sim {1\over |\xx-\yy|^{2 X_{CR} }}\label{is1s}\nn
\eea
Another planar lattice model is the {\it dimer} model, whose
partition function is
\be Z_\L=\sum_{M\in \mathcal M_{\L}}
\Big[\prod_{b\in M}t_b\Big]\label{cc}
\ee
where $M$ is the set of dimer coverings (or perfect matchings) of $\L$, that is
a subset of edges such that each vertex
of $\L$ is contained in exactly one edge in $M$. The lattice is bipartite (black and white sites) 
and the weights $t_1,t_2,t_3,t_4$ are associated to edges
with white endpoint to the
right, above, to the left or below the black endpoint.

The {\it interacting } dimer model has partition function given by
\be Z_\L=\sum_{M\in \mathcal M_{\L}}
\Big[\prod_{b\in M}t_b\Big]
 e^{\l \sum_\xx f(\t_\xx M) }\label{cc13}
\ee
where $f$ is a local function around the origin, $\t_x$ translates by $x$. In a suitable region of the weights (liquid phase) one expects
that the {\it dimer-dimer} correlations decay as power law and have the form
\be \media{I_b;I_{b'}}=
(-1)^{x_1+x_2}\frac{A}{2\p^2}{\rm Re}\frac1{z^{2}}
+(-1)^{x_1}\frac{K_2(\l)}{|z|^{2\h_1}}+h.o.\label{2222}
\ee
with $b=(\xx,\xx+e_i)$, $b'=(\yy,\yy+e_i)$, $z=(x_0-y_0)+i (x_1-y_1)$.

The {\it height} difference between $\x$ and $\h$ is
$$ 
h_\x-h_\h=\sum_{b\in \mathcal C_{\x\to \h}}
(I_b(M)- {1\over 4} )\s_b
$$
where $\x,\h$ label faces,  $\s_b=+1/-1$ depending on whether $C_{\x\to \h}$ 
crosses $b$ with the white site on the right/left;  $I_b(M)$ is equal to 1 if $b$ is occupied by a dimer
in $M$, and 0 otherwise. 
$h_\x-h_{\h}$ is {\it independent} of the choice of
$C_{\x\to \h}$. 

In the special case of plaquette interaction the interacting dimer model
is equivalent to {\it six vertex} model,
provided that the 6V has weights $a_1,..,a_6$,
\be
t_1=a_1,\quad\;t_2=a_4,\quad\;t_3=a_2,\quad \;(t_1 t_3+t_2)e^\l=a_6\label{sss}
\ee
With the above choice of parameters, there is a relation between dimer model and Ashkin-Teller exponents with a suitable choice of the parameters, as shown in \cite{D}; 
\be
X_e=\h_1  \quad \quad X_P=X_A\label{sap}
\ee
where $X_e, X_P$ are the Ashkin-Teller exponents \pref{is1s}, $\h_1$
is the exponent appearing in the dimer correlation \pref{2222} and $X_A$ is the exponent of the electric correlation
\be
\media{e^{i\pi(h_\xx-h_\yy)}}\sim |\xx-\yy|^{-2X_A}
\ee
%

\section{Exact solutions and universality conjectures}

The celebrated Onsager solution \cite{O} of the nearest-neighbor Ising model 
\pref{is} gives the  exact values of the critical temperature $\tanh\b_c J=\sqrt{2}-1$ and of the critical exponents
\be
\h=1/4\quad\quad \n=1\quad\quad X_e=1
\ee
In the above expression already a phenomenon of universality appear; 
the only "microscopic" parameter is $J$ and, while the critical temperature depends on it,
the exponents are independent.

Similarly in the case of coupled Ising models \pref{c11} the model is solvable at $\l=0$; there are in general two
critical temperatures $\tanh\b_c^1 J_1=\sqrt{2}-1$ and $\tanh\b_c^2 J_2=\sqrt{2}-1$ which coincide at $J_1=J_2$ and
\be
X_{P}=1/4 \quad\quad X_{e}=X_{C R}=1
\ee
In general \pref{c11} is not solvable at $\l\not=0$ except that in the special case 
\pref{111} when it reduces to the eight-vertex model; this model was solved by Baxter and one critical exponent can be computed
\be
\n={\pi\over 2\m}\quad\quad \tan (\m/2)=(cd/ab)^{1\over 2}=e^{-4\l}\label{pik}
\ee
Note that the exponent depend on the microscopic parameters of the Hamiltonian
when $\l\not=0$.

The dimer model \pref{cc} was solved by Kasteleyn \cite{Ka}
in the case $t_b=1$ and the expression 
\pref{2222} was proved with
\be
\h_1=1  \quad\quad A=1
\ee
Moreover in \cite{kos} it has been proved that
the height field is asymptotically a free gaussian field for values of the weights in the
liquid phase
\be
\media{(h_\x-h_{\h})^2}=\frac{A}{\p^2}\log|\x-\h|+R(\x-\h)\quad\quad A=1\label{2223}
\ee
with $R$ bounded; in addition the truncated expectations are bounded. 
Note that the prefactor $A=1$ is independent from the weights $t_i$. This is another manifestation of universality; it regards the amplitude 
of the dimer correlations
but it is also related to the universality of one exponent. Indeed by \pref{pik} the electric exponents is
\be X_A=A/4\label{sap1}  \ee
as 
\be
\media{e^{i\pi(h_x-h_y)}}\sim
e^{-\frac{\pi^2}{2}\media{(h_x-h_y)^2}}\sim
e^{-\frac{A}2\log|x-y|}\label{6V} 
\ee

All the above results are based on {\it exact solutions} and we can distinguish two kinds of them. The first are the so called
{\it free fermion} solutions; in such cases the partition function can be written as a Pfaffian or a determinant. The name is due to the fact that 
determinants can be represented in terms of Gaussian Grassmann integrals, which in quantum physics are used to describe fermionic systems
without interaction. It belongs to this class the nearest neighbor Ising model or the non-interacting dimer model.
In the case of coupled Ising model \pref{c11} or generalized dimer models \pref{cc13}
the determinantal representation is not valid and the representation is in terms of non gaussian Grassmann integrals.
There are very few cases, like the eight or the six vertex model, where a solution can be found, but generically  no solution exists. It is also important to distinguish between 
local and extended exponents. Local correlations are  
the energy or crossover correlations in Ising models,
and dimer correlations in dimer models; extended ones are the height or electric correlations in
dimer models or the spin-spin correlation. Note that indeed the spin correlation in Ising 
models are rather similar to the electric correlations in dimer models, see \cite{MW}.

The exponents discussed above are all computed by exact solutions. What can we do when a solution is lacking?
It was conjectured that universality holds for the generalized Ising model \pref{c7}, in the sense that the exponents should coincide with their $\l=0$ value, which coincides with the solvable nearest neighbor model. However there are counterexamples to a  too strong notion of universality.
The coupled Ising model \pref{c11} reduces to the Ising model as $\l=0$;
in this case however the exponents depend on $\l$, as it appears from \pref{pik}.
For models with continuously varying exponents one can define a weaker form of universality.
It was proposed by Kadanoff \cite{Ka} that all the critical exponents can be determined by the
knowledge of a single one, and in particular the following relations hold
\be
X_{CR}(\l)=1/X_e(\l) \quad\quad X_P(\l)=X_e(\l)/4\quad\quad \n={1\over 2-X_e(\l) }\label{cv}
\ee
Even if the exponents depend on all the microscopic details, the knowledge of one on them allow the exact prediction of the others. Some check of the above relations can be obtained
in special solvable cases using a mapping in the $XXZ$ spin chain, see \cite{Ha}, but of course
one expects that the validity of such relations holds both in solvable and non solvable cases. 

It is believed that to this class of universality belongs the Gaussian Free fields $\Phi$
with covariance 
\be
E(\Phi_\xx\Phi_\yy)=-A/2\pi^2 \log|\xx-\yy|
\ee
with $A\not=1$, while the case $A=1$ correspond to the Ising class. Indeed 
by a suitable identification of the correlations of coupled Ising or dimers with the ones of
functions of derivatives or exponentials  of $\Phi$
one gets exponents all parametrized by $A$ and verifying the relation \pref{cv}.
A gaussian free field appears indeed formally starting from the Grassmann representation and performing a scaling limit, see
\cite{K11},\cite{K12},\cite{K13}; the limit identifies a Quantum Field Theory of interacting fermions which can be mapped in a gaussian theory, a fact known as
{\it bosonization}.
Converting this argument in a rigorous derivation
of the scaling relations is however possible only in the nearest neighbor Ising or dimer model, see  \cite{D}, \cite{kos}.


\section{Rigorous Renormalization Group}

The universality conjecture has a natural interpretation 
in the picture given by the Renormalization Group. In  the case of the generalized Ising model \pref{c7}
 one expects that the system has a trivial fixed point corresponding to the nearest-neighbor Ising model, while in the 
case of coupled Ising models \pref{c11} there should exist a line of fixed points. 
The most intuitive approach would be
to define an RG transformation directly in the spin variables, but this encounters
a number of difficulties. A more convenient approach is to use  
the fact that the exact solution of the 
nearest-neighbor Ising model 
can be written as sum of Pfaffians
of suitable matrices. The exact solvability relies on the fact that such Pfaffians can be explicitly computed.
The Pfaffian can be formally written as Grassmann gaussian integrals.
If $\psi_i$ are a finite set of Grassmann variables
and $A$ is a $2n\times 2n$ antisyimmetric matrix, 
\be
Pf A= \int \prod_{i=1}^{2n} d\psi_i e^{ -{1\over 2} (\psi, A\psi)}
\ee
with $\int \prod_{i=1}^{2n} d\psi_i  \psi_1...\psi_{2n}=1$ and zero otherwise.
%
%

Given an anti-symmetric matrix $M$ we define the Gaussian
Grassmann measure with “propagator” M, denoted $P_M(d\psi)$, which maps a
polynomial $f$ of the $\psi$ variables into a complex number denoted
by $\int P_M(d\psi) f(\psi)$ or $<f>$; the map
is linear, $<1>=1$ and \be <\psi_{k_1}...\psi_{k_n}>=Pf (M(k_i,k_j))_{i,j\le M}\ee
If $M$ is invertible we can write more explicitly as
\be
<f>={\int \prod_{i=1}^{2n} d\psi_i e^{ -{1\over 2} (\psi, M^{-1} \psi)} f(\psi)\over 
\int \prod_{i=1}^{2n} d\psi_i e^{ -{1\over 2} (\psi, M^{-1}\psi)}   }=\int P_M(d\psi) f(\psi)
\ee
Of course $P_M(d\psi)$ is not a measure in the usual probabilistic sense. 

It turns out that the partition function of \pref{c7} or \pref{c11}
can be expressed in terms of
Grassmann integrals of the form
\be
\int P_g(d\psi) e^V\label{c16aa}
\ee
where $V$ is sum of monomials in the $\psi$ of degree more than fore and $g$
is a suitable covariance decaying exponentially at temperatures different from the critical temperature  $\b\not=\b_c$ and as a power with exponent $1$
at the critical temperature, $\tanh \b_c J=\sqrt{2}-1$. When $\l=0$ then $V=0$ and the partition
function reduces to a gaussian Grassmann integral corresponding to the Ising model.

The Grassmann integral \pref{c16} has a strong resemblance with the $\phi^4$ model, which
is also strongly related to the Ising model, that is 
\be
\int P(d\phi) e^{-\l\sum_\xx \phi^4_\xx}\label{c16a}
\ee
where now $\phi_\xx$ range over all $\RR$ and $P(d\phi)$ is a gaussian measure.
Despite their formal similarity, there are deep differences between the above two expressions;
actually the second require $\l>0$ so there is no hope for analyticity in $\l$, while 
there is no restriction on the sign of $\l$ in the first, as at finite $L$ is simply a polynomial in 
$\l$. It should also remarked that \pref{c16a} reduces at $\l=0$ to mean field, while
the first, valid only in $d=2$, it reduces to the Ising model solution; this should make clear why
the Grassmann representation is more convenient in two dimension.

The (rigorous) Renormalization Group analysis of \pref{c16aa} is based
on the addition property, saying that, if $g=g_1+g_2$  
\be
\int P_g(d\psi) f(\psi)=\int P_{g_1}(d\psi_!)\int P_{g_2}(d\psi_1) f(\psi_1+\psi_2)
\label{d1}
\ee
Moreover 
\be
\int P_g(d\psi) e^{V(\psi+\psi_1)}=e^{\sum_{n=0}^\io {1\over n!} \EE_\psi^T(V;n)}\equiv
e^{\bar V(\psi_1)}\label{d2}
\ee
where  $\EE_\psi^T(V;n)$ are the {\it truncated expectations} or cumulants. One writes
the covariance $g$ as
\be
g(\xx,\yy)=\sum_{h=-\io}^0 g^{(h)}(\xx,\yy)\label{d3}
\ee
with $g^{(h)}(\xx,\yy)$ of size $\g^h$ and decaying faster than any power with rate $\g^h$,
where $\g>1$ is a scaling parameter. By using \pref{d1}, \pref{d2}, \pref{d3}
we can write, if $g=g^{(\le -1)}+g^{(0)}$, $g^{(\le -1)}=\sum_{h=-\io}^{-1} g^{(h)}$
\bea
&&\int P_g(d\psi) e^{V(\psi)}=\int P_{g^{(\le -1)}}(d\psi^{(\le -1)}) \int P_{g^{(0)}}(d\psi^{(0)}) 
e^{V(\psi^{(\le -1)}+\psi^{(0)}
)}=\nn\\
&&\int P_{g^{(\le -1)}}(d\psi^{(\le -1)}) 
e^{V^0(\psi^{(\le -1)})}
\label{c16}
\eea
After the integration of $\psi^{(0)}$ one gets an expression similar to the initial one,
except that the gaussian integration has covariance $g^{(\le -1)}$ instead of $g$,
$V^0(\psi^{(\le -1)})$ is sum of monomials in the $\psi^{(\le -1)}$ summed over suitable kernels $W^{(-1)}$. Such kernels are, in the infinite volume limt, expressed by series in $\l$
which are uniformly convergent so that $W^{(-1)}$ is analytic in $\l$ in a disk around the origin.
The proof of this fact follows from the Brydges-Battle-Federbush \cite{Br}
representation of the truncated expectations in terms 
of sum of pfaffians, which can be bounded by the Gram-Hadamard inequality. 
One iterates the above procedure getting
\be
\int P_g(d\psi) e^{V(\psi)}=\int P_{g^{(\le h)}}(d\psi^{(\le -1)}) 
e^{V^h(\psi^{(\le h)})}
\ee
with $V^h(\psi^{(\le h)})$ sum of monomials in $\psi^{(\le h)}$ times suitable kernels 
$W^{(h)}$. One associates a {\it scaling 
dimension} to the monomials of order $n$.
This notion is crucial in the above analysis. 
By applying the Brydges-Battle-Federbush formula and Gram bounds one 
still get analyticity in $\l$ for such kernels, but the radius of convergence shrinks to zero
unless $D_n<0$ for any $n>2$; the case $n=2$ is somewhat special 
as quadratic terms can be inserted in the gaussian measure and produce a shift in the critical temperature. It turns out that the dimension is the same for both \pref{c7} or \pref{c11},
namely 
\be D_n=2-n/2-s\ee  
where $s$ is the order of (discrete) derivatives
and that $D$ is negative except that for $n=4, s=0$. In the standard RG terminology, the terms with negative dimension are called irrelevant, the ones with vanishing dimension are marginal and the ones with positive dimension are the relevant.
Up to this point the analysis for  
\pref{c7} or \pref{c11} shows similar features; here however comes the crucial difference.

In the case of \pref{c7} to each $\xx$  are associated two Grassmann variables, actually $\psi_\xx, \bar\psi_\xx$; in the case of \pref{c11}, corresponding to two Ising models, the number of Grassmann variables is twice, actually $\psi^1_\xx, \bar\psi^1_\xx,\psi^2_\xx, \bar\psi^2_\xx$.
as a consequence, in the case of \pref{c7} there are no terms $n=4, s=0$
as, by the anticommutativity of the Grassmann variables
\be
\psi_\xx\bar\psi_\xx\psi_\xx\bar\psi_\xx=0
\ee
The quartic terms must contain necessarily variables with different coordinates, and by anticommutativity they can be written as sum of terms with at least a derivative,
hence $D_n<0$.
Therefore the kernels remain analytic uniformly in $h$ and, as a consequence of the irrelevance of all the terms in the effective potential with $n\ge 4$, the correlations are equal to the non interacting ones plus terms decaying in the same way; that is, the exponents do not change with 
respect to the Ising model ones. The following theorem holds.
\vskip.3cm
{\bf Theorem}{\it
(Pinson-Spencer \cite{Sp})  In the generalized Ising model \pref{c7}
for $\l$ small enough  there is a critical temperature $\tanh J\b_c(\l)=\sqrt{2}-1+F(\l)$ with $F(\l)=a\l+O(\l^2)$
such that the energy correlation decay as in \pref{sap}
with exponents
\be X_e(\l)=1\quad\quad \n(\l)=1\ee
} 
\vskip.3cm
The above Theorem says that universality  indeed holds for the generalized Ising model; the $X_e,\n$ exponents are the same as the Ising one, while the critical
temperature is different.
\vskip.2cm
Let us consider now the case of coupled Ising models 
\pref{c11} with $J'=J$. Now there are marginal terms
of the form $\psi^1_\xx\bar\psi^1_\xx\psi^2_\xx\bar\psi^2_\xx$; their presence has the effect that the estimated radius of convergence shrinks to zero as $-h\to\io$. One has therefore to 
modify the integration procedure so that $V^h$ has the form
\be
V^h=\l_h \sum_\xx \psi^1_\xx\bar\psi^1_\xx\psi^2_\xx\bar\psi^2_\xx+ \RR \VV^h
\label{bb1}
\ee
where $\RR \VV^h$ contains now only monomials with negative dimension; moreover the
integration is with respect to a variance of the form $g^h/Z_h$. With
this modified procedure the kernels in  $\RR \VV^h$ becomes now functions of $\l_h$,
and analytic in $\l_h$ belonging to a disk around the origin. Both $\l_h$ and $Z_h$ verify finite difference equations of the form
\be
\l_{h-1}=\l_h+\b_\l^h(\l_h,...,\l_0)\quad \quad  {Z_{h-1}\over Z_h}=1+
\b_z^h(\l_h,...,\l_0)
\ee 
with $\b_\l^h,\b_z^h$ expressed by convergent series in $\l_k$. It turns out, by an explicit computation,that 
\be
\b_\l^h(\l_h,...,\l_0)=a\l_h^2+O(\l_h^3)\quad
\b_z^h=b \l_h^2+O(\l_h^3)
\ee 
with 
\be
a=O(\g^h)\quad\quad
b>0 
\ee
with $b$ an $h$-independent constant.
If $a<0$ and $h$ independent, we could conclude that $\l_h$ decrease with $h$, 
and if $a>0$ the opposite conclusion could be taken. 
Therefore the result of a second order perturbative computation is sufficient
to determine the behavior of $\l_h$.  In the present case however this is not sufficient;
terms $O(\g^h)$ can never dominate over the rest, unless a cancellation is proven at any order. In this sense the proof of the exisitence of a line of fixed points is a non-perturbative problem. One needs to prove that $\b_\l^h$  is asymptotically vanishing,
a fact established in \cite{BM}
\be
|\b_\l^h(\l_h,...,\l_h)|\le C \g^h \l_h^2 \label{van}
\ee
The idea of the proof will be briefly recalled below. From this expression we get that
$\l_h=\l+O(\l^2)$ 
\be
\lim_{h\to-\io}\l_h=\l_{-\io}(\l)
\ee
with $\l_{-\io}(\l)$ an analytic function of $\l$. Therefore \pref{van} says that there is a line of fixed points. Moreover
\be
\lim_{h\to-\io} {Z_{h-1}\over Z_h}=\g^\h
\ee
with $\h=\h(\l_{-\io})$ analytic function of $\l_{-\io}$ and therefore of $\l$.
Other critical exponents are associated to the correlation length and to the energy and crossover
correlations. The result can be summarized by the following result.
\vskip.3cm
{\bf Theorem}{\it
(
Mastropietro \cite{BFM}).
In coupled Ising models with Hamiltonian \pref{c11}  $J=J'$ and small $\l$ 
at $\b=\b_c(\l)$ the correlations
decay as in \pref{is1s} with
energy $X_e(\l)$ and crossover $X_{CR}(\l)$ exponents analytic functions of $\l$ given by, $a\not= 0$ constant 
\be
X_e(\l)=1- a \l+O(\l^2) \quad X_{CR}(\l)=1+a \l+O(\l^2) \quad \nu(\l)=1- a \l+O(\l^2) 
\ee
}.
\vskip.2cm
The exponents are expressed by convergent series in $\l$, with coefficients depending on all the microscopic details. This allows to compute the exponents 
with arbitrary precision and to conclude the existence of a line of fixed points.

Finally when $J\not =J'$ in coupled Ising models with Hamiltonian \pref{c11} 
there are two critical temperatures and the exponents coincide with the Ising ones; a critical exponent appear instead in the difference of critical temperatures
\cite{GM} 
\be
|\b^1_c(\l)-\b^2_c(\l)|=O(|J-J'|^\m)\ee
with $\m=1+c\l+O(\l^2)$.

\section{The effective QFT model and universality}

The above analysis shows that the exponents can be written as convergent series in $\l$, and this allows to compute them
with arbitrary precision, by explicitly computing lowest order with a rigorous bound of the rest.
However the series are so complicate that an explicit verifications of the universal relations is essentially impossible from them.
In order to prove universality, and to prove \pref{van}, we introduce a reference model which is related to the formal continuum limit
of the lattice model.
The partition function of this effective model is 
the following Grassmann integral (similar definition holds for the generating function)                                                             
\be
\int  P(d\psi^{(\le N)}) e^{
\tilde\l  Z^2 \int d\xx d\yy v(\xx-\yy) \r_{+,\xx}\r_{-,\yy}
+\sum_\o
Z^{(1)} \int d\xx J^1_{\o,\xx} \r_{\o,\xx}
}\label{ddd} 
\ee
with $\xx=(x_0,x_1)$ with $\xx\in \L\subset \RRR^2$ (that is the model is defined on the continuum and not on a lattice), $\o=\pm 1$,
$P(d\psi)$ is the fermionic integration with propagator $\hat g_\o(\kk)={1\over Z}{\chi_{N}(\kk)\over D_\o(\kk)
}$ with $D_\o(\kk)=-i k_0+\o v k_1$, $\chi_N(\kk)$ is a smooth cut-off function vanishing for
$|\kk|\ge \g^N$, $v(\xx-\yy)$ a short range symmetric potential and finally $\r_{\o,\xx}=\psi^+_{\o,\xx}\psi^-_{\o,\xx}$.

The model can be analyzed by RG and the analysis 
is similar to the one for coupled Ising models 
\pref{c11} . The crucial observation is that the propagator at scale $h$ differs by terms $O(\g^h)$ (so smaller and smaller as $-h$ increases)
and the interaction  differs by irrelevant terms with negative dimension.
The result is that 
the scaling dimension is the same,
the effective potential has the form \pref{bb1}
and 
the beta function $\b_\l^h$ is essentially identical to the lattice beta function 
for \pref{c11} 
up to $O(\g^h)$ terms. Of course the bounds for the kerneles $W^h$ are the same but not their values, as their propagators and the interaction are different. It should also be remarked that this is true for scales $h\le 0$; one needs in this case a multiscale analysis
also for the positive scales, but it can be performed due to the non-locality of the interaction.

We have then introduced a model which has asymptotically the same beta function as the coupled Ising seen above. The advantage is however that this model as $N\to\io$ is solvable,
in the sense that the correlations obey to closed equations.
This is due to the fact that the model \pref{ddd}
has extra symmetries which are only approximate for 
coupled Ising models  \pref{c11}.
These symmetries imply
identities , called {\it Ward Identities}, which are obtained performing the change of variables $\psi^\pm_\o\to e^{\pm i \a_{o,\xx} }\psi^\pm_{\o,\xx}$
in \pref{ddd}. One obtains  ($<>$ are the derivatives with respect to $J$)
\be
D_\o <\hat\r_{\pp,\o}\hat\psi^+_{\kk,\o'}\hat\psi^-_{\kk+\pp,\o'}>+
\D_N(\kk,\pp)=
\d_{\o,\o'}  {Z^{(1)}\over Z} [<\hat \psi^+_{\kk,\o'} \hat \psi^-_{\kk,\o'}>  - <\hat\psi^+_{\kk+\pp,\o'} 
\hat\psi^-_{\kk+\pp,\o'}>\label{lll}
\ee
where $\D_N=<\d\hat\r_{\pp,\o}\hat\psi^+_{\kk,\o'}\hat\psi^-_{\kk+\pp,\o'}
>$ with 
\be
\d\hat\r_{\pp,\o}=
\int d\kk
[(\chi^{-1}_N(\kk+\pp)-1)D_\o(\kk+\pp)
-(\chi^{-1}_N(\kk)-1)D_\o(\kk)]\hat\psi^+_{\kk,\o}\hat\psi^-_{\kk+\pp,\o}\ee
The $\D_N$ term is not vanishing even if $N\to\io$ a fact known as {\it chiral anomaly} \cite{AB}.
In the limit one gets
\be
\lim_{N\to\io} \D_N(\kk,\pp)=\t\hat v(\pp)  D_{-\o}(\pp)
<\hat\r_{\pp,-\o}\hat\psi^+_{\kk,\o'}\hat\psi^-_{\kk+\pp,\o'}>\label{sapp}
\ee 
with
\be \t={\tilde\l \over 4\pi v}\ee
By inserting this expression in \pref{lll} one gets a {\it closed} equation between correlations. In order to use such identities for controlling $\tilde\l_h$ one consider a sequence of models with infrared cut-off $\g^h$ and write the closed equation for the four-point correlation which are proportional to $\l_h$ at vanishing moments; 
from the closed equation one gets
\cite{BM}
\be
\tilde\l_h=\tilde\l+O(\tilde\l^2)
\ee
and this is possible only if \pref{van} holds. As the beta functions is the same for the 
coupled Ising model, this reult can be used to prove the existence of a line of fixed points also in the coupled Ising model.

The universal relations followss from another crucial property of \pref{sapp}, actually that $\t$ is exactly linear in $\tilde\l$.
This is a rather remarkable property; in general all expression we have encountered are series in the coupling, but only the anomaly has no higher order contributions, a fact known as 
{\it non-renormalization} of the anomaly. Another crucial step is that we can choose
$\tilde\l$ in the reference model as function of $\l$ so that
so that $\tilde\l_{-\io}(\tilde\l)= \l_{-\io}(\l)$. As the exponents are function of $\l_{-\io}$, they therefore coincide in the reference and lattice model.
On the other hand using \pref{lll} combined with
other identities named Schwinger-Dyson equation one gets a closed expression for the correlations and 
exact expressions for the exponents
\be X_e=
{1-\frac{\tilde\l}{4\pi v} \over 1+\frac{\tilde\l}{4\p v} } \quad X_{CR}=
{1+\frac{\tilde\l}{4\pi v}\over 1-\frac{\tilde\l}{4\p v} }
\ee
Note that $\tilde\l$ is a complicate expression in $\l$ but the exponents have a simple expression in terms
 of $\tilde\l$ allowing the verification of the universality relations.  
In conclusion one gets the following result.
\vskip.3cm
{\bf Theorem}(Benfatto Falco Mastropietro \cite{BFM},\cite{M2})
{\it
In coupled Ising models \pref{c11}
 and small $\l$ the following relations are true
$$X_e(\l)  X_{CR}(\l)=1 \quad\quad \n={1\over 2-X_e(\l)}\quad\quad \m={2-X_e(\l)\over 2-X_{CR}(\l)} $$
}.

This proves universality; even if exponents depend on all microscopic details, one can express all in terms of a single one of them.
Note also that it is sufficient the solvability of the reference continuum model
to get results for the non solvable lattice ones.

\section{Interacting dimers}

The above universality results hold for exponents corresponding to local quantities,
that is expressed by monomials in the Grassmann variables $\psi_\xx$ with the same coordinate. 
Relations for exponents associated to non-local
expressions have been more recently derived for the dimer model \pref{cc13}.
It is possible to repeat the RG analysis in the present case and
prove that the dimer correlation has the form \pref{2222} with 
$
\h_1=1+a\l+(\l^2)
$.

By using that expression \pref{2222}
one can extend the analysis in \cite{kos} to prove that the height function is asymptotically a Gaussian free field even in the interacting dimer model. One can write
variance as
$\langle\,(h_{\xx}-h_{\yy})^2\,\rangle$ as the sum of two terms.
In the first  
term the sign  {\it compensates exactly} the
sign coming from the factor $\s_b$ and one gets 
\be -\frac{A(\l)}{2\pi^2}\sum_{b_1\in \mathcal C^{(1)}_{\xx\to \yy},
b_2\in \mathcal C^{(2)}_{\xx\to \yy}}{\rm Re}\frac{\D z_{b_1}\D z_{b_2}}{(z_{\xx_1}-z_{\xx_2})^2}\;,\label{5.95}\ee
where $z_{b_1},z_{b_2}$ are points on the two paths and
$\D z_{b_i}$ is the displacement associated with the elementary portion of the path 
$\mathcal C^{(i)}$ crossing $b_i$, thought of as a complex vector of
modulus 1. Such term is
the Riemann approximation to the integral
\be -\frac{A(\l)}{2\pi^2}{\rm
  Re}\int_{z_{\xx}}^{z_{\yy}}dz\int_{z_{\xx}}^{z_{\yy}}dw\frac1{(z-w)^2}\label{4.int}\ee
Note the crucial point that, thanks to the independence from the
path, the two paths can be chosen far away. It is also important to stress that
no anomalous exponent is found in the first term of \pref{2222}.
Regarding the contribution with exponent $2\h_1$, one notices that
the oscillatory $\s_{b_1}\s_{b_2}$ {\it do not compensate} the
oscillatory factor $(-1)^{(\xx_1-\xx_2)_{j_1}}$.  Once summed over the
paths, 
the oscillatory factor
has the same effect {\it as a discrete
derivative}; therefore such term produces an $O(1)$ contribution (instead than logarithmic) just
for dimensional reasons.  In addition, a universal relation is found for the amplitude $A(\l)$, as it appears by the following Theorem.
\vskip.3cm
{\bf Theorem} (Giuliani Mastropietro Toninelli \cite{GMT1}, \cite{GMT2}
){\it For $\l$ small enough the dimer correlation has the form
\pref{2222} with $
\h_1=1+a\l+(\l^2)$ and the dimer correlation verifies \pref{2223} with
\be
A(\l)=\h_1(\l)\label{2224} 
\ee
}
\vskip.3cm
Higher  cumulants of the height difference between two points are bounded uniformly in their distance. As a conclusion, the height function
converge to the massless Gaussian field 
also in the interacting case, the only difference being that the amplitude
is $\l$-dependent. It turns out the universality holds, in the sense
that the amplitude is exactly equal to the critical exponent in the dimer correlation, see \pref{2224}; such result is exactly what one expects if the picture provided by bosonization is true.
In addition, by \pref{sap} and \pref{sap1} we get
\be
X_e=\h_1=A=4 X_A=4 X_P
\ee
which is the Kadanoff relation for the Ashkin-Teller model; the universality
conjecture is therefore established also for extended exponents like the polarization, at least in the case of Ashkin-Teller model.

In order to prove \pref{2224} we still consider the reference model and by \pref{lll}  we get the density correlations ($v=1$ in the dimer model)
\be
\media{\hat \r_{p,\o};\hat\r_{-p,\o}}= -\frac{1}{Z^2}\frac{(Z^{(1)})^2}{1-\frac{\l_\infty^2}{16\pi^2}}
\frac{D_{-\o}(p)}{D_\o(p)}.\ee
which says that $A$ appearing in the dimer corrrelations is
\be
A(\l)=\frac{(Z^{(1)})^2}{Z^2(1-\frac{\l_\infty^2}{16\pi^2})}.\ee
Note that $Z^{(1)}$ and $Z$ are chosen as series in $\l$ so that the corresponding effective couplings in the dimer and effective model are the same.
By comparing \pref{lll}
with an analogue identity derived directly in the lattice dimer model we get the relation
\be
\frac{Z^{(1)}}{(1+\frac{\l_\infty}{4\pi})Z}=1
\ee
Moreover in the effective model
\be
\h_1=\frac{1+\frac{\l_\infty}{4\pi}}{1-\frac{\l_\infty}{4\pi}}\ee
and this implies that 
\be
A(\l)=\frac{1+\frac{\l_\infty}{4\pi}}{1-\frac{\l_\infty}{4\pi}}=\h_1.\ee

\section{Conclusions}

We have reviewed the proof of universality 
in the case of line of fixed points for planar lattice models including coupled Ising,
vertex and dimer models.
Open  problems include the relations for other exponents 
like the spin ones, 
more general boundaries and optimal estimates on the size of the coupling.
\vskip.3cm
{\bf Acknowledgements.} 
This article is based on the talk at the mathematical physics workshop “Inhomogeneous Random
Systems” (Institut Curie, Paris, January 28, 2020). The work has been supported by MIUR, PRIN 
2017 project MaQuMA. PRIN201719VMAST01.


\begin{thebibliography}{999999}



\bibitem{B1}
C.H. Back, Ch. W¨ursch, A. Vateriaus, U. Ramsperger, U. Maier, D.
Pescia. Experimental confirmation of universality for a phase transition
in two dimensions. Nature 378 (1995) 597-600.
\bibitem{B2}
J. C. Campuzano, M. S. Foster, G. Jennings, R. F. Willis, W. Unertl,
Au(110) (1×2)-to-(1×1) Phase Transition: A Physical Realization of the
Two-Dimensional Ising Model.  Phys. Rev. Lett. 54, 2684-2687 (1985).
\bibitem{O}
L. Onsager. Crystal Statistics. I. A Two-Dimensional Model with an
Order-Disorder Transition. Phys. Rev. 65, 117-149 (1944).
\bibitem{Ba}R.J. Baxter. Exactly solved models in statistical mechanics, Academic Press, London, (1989).
\bibitem{B3} N. Khan, P. Sarkar, A. Midya, P. Mandal  P. K. Mohanty. 
Continuously Varying Critical Exponents Beyond Weak Universality.
Scientific Reports volume 7, Article number: 45004 (2017).
\bibitem{Ka}
Kadanoff L.P. Connections between the Critical Behavior of the Planar Model
and That of the Eight-Vertex Model. Phys. Rev. Lett. 39, 903–905, (1977).
\bibitem{Ha}
F. D. M. Haldane, General Relation of Correlation Exponents and Spectral
Properties of One-Dimensional Fermi Systems: Application to the Anisotropic
S = 1/2 Heisenberg Chain.  Phys. Rev. Lett. 45, 1358-1362 (1980)
\bibitem{B4}
 R. R. Nair, P. Blake, A. N. Grigorenko, K. S. Novoselov,
T. J. Booth, T. Stauber, N. M. R. Peres, A. K. Geim,
Science 320, 1308 (2008).
\bibitem{W}
K. G. Wilson and J. Kogut. The renormalization group and the $\e$
expansion. Physics Reports 12, 75–199 (1974).
\bibitem{LP} Luther A., Peschel I.: Calculations of critical exponents in two dimension from
quantum field theory in one dimension. Phys. Rev. B 12, 3908–3917, (1975).
\bibitem{K11}
L.P. Kadanoff, A.C. Brown. Ann. Phys. 121, 318–345 (1979).
\bibitem{K12}
den Nijs M.P.M.: Derivation of extended scaling relations between critical exponents in two dimensional models from the one dimensional Luttinger model.
Phys. Rev. B 23, 6111–6125, (1981).
\bibitem{K13}
Pruisken A.M.M. Brown A.C.: Universality for the critical lines of the eight
vertex, Ashkin-Teller and Gaussian models. Phys. Rev. B 23, 1459–1468, (1981).
\bibitem{D}
J. Dubedat: Exact bosonization of the Ising model, arXiv:1112.4399
\bibitem{kos}
R. Kenyon, A. Okounkov, S. Sheffield: Dimers and amoebae. Ann. Math. 163,
1019-1056 (2006).
\bibitem{M1}
V. Mastropietro: Ising models with four spin interaction at criticality, Comm.
Math. Phys. 244, 595-642 (2004).
\bibitem{BM}
G. Benfatto, V. Mastropietro: Ward identities and chiral anomaly in the Luttinger liquid, Comm. Math. Phys. 258, 609-655 (2005).
\bibitem{GM}
A. Giuliani and V. Mastropietro. Anomalous universality in the
anisotropic Ashkin–Teller model. Communications in Mathematical
Physics 256, 681–735 (2005).
\bibitem{BFM} G. Benfatto, P. Falco, V. Mastropietro: Extended Scaling Relations for Planar
Lattice Models. Comm. Math. Phys. 292, 569-605 (2009).
\bibitem{M2} V.Mastropietro Universality, Phase Transitions and Extended Scaling Relations. Proceedings of the International Congress of Mathematicians 2010 (ICM 2010), pp. 2078-2104 (2011)
\bibitem{GMT1}  A. Giuliani, V. Mastropietro, F. Toninelli. Height fluctuations in interacting
dimers, Ann. Inst. Henri Poincare' (Prob. Stat) 53, 98-168 (2017)
\bibitem{GMT2} A. Giuliani, V. Mastropietro, F. Toninelli
Non-integrable Dimers: Universal Fluctuations of Tilted Height Profiles.
Commun. Math. Phys. 377, 1883–1959 (2020)
\bibitem{BFM3} 
G. Benfatto, V. Mastropietro, P.Falco.
Universality of One-Dimensional Fermi Systems, I. Response Functions and Critical Exponents.
Communications in Mathematical Physics volume 330, 153–215(2014) 
\bibitem{BFM4} 
G. Benfatto, V. Mastropietro, P.Falco
Universality of One-Dimensional Fermi Systems, II. The Luttinger Liquid Structure
Communications in Mathematical Physics volume 330, 217–282 (2014)
\bibitem{AMP}G. Antinucci, V. Mastropietro, M. Porta. 
Universal Edge Transport in Interacting Hall Systems. 
Communications in Mathematical Physics volume 362,  295–359 (2018)
\bibitem{MW}
B. McCoy and T. Wu. The two-dimensional Ising model. Harvard
University Press (1973)
\bibitem{Kas}
P. Kasteleyn, Graph theory and crystal physics, in: “Graph Theory and Theoretical Physics”, pp. 43-110 Academic Press, London (1972)
\bibitem{Br}
D. Brydges: A short course on cluster expansions, in Phenomenes critiques,
syst`emes al´eatoires, th´eories de jauge, Les Houches summer school session 43,
pages 129-183, North-Holland (1986).
\bibitem{Sp}
T. Spencer. A mathematical approach to universality in two dimensions. Physica A Statistical Mechanics and its Applications 279,
250–259 (2000)
\bibitem{AB} S. L. Adler, W. A. Bardeen.  Absence of Higher-Order Corrections in the Anomalous Axial-Vector Divergence Equation
Phys. Rev. 182, 1517 (1969).
\end{thebibliography}
\end{document}